\documentclass[a4paper,11pt]{article}
\usepackage{pos}
\usepackage{graphicx}

\title{Towards the phase diagram of cold and dense heavy QCD}
\ShortTitle{Towards the phase diagram of cold and dense heavy QCD}

\author*{Amine Chabane}
\author{Owe Philipsen}

\affiliation[a]{Institute for Theoretical Physics, Goethe University, Max-von-Laue-Strasse 1, 60438 Frankfurt}

\emailAdd{chabane@itp.uni-frankfurt.de}
\emailAdd{philipsen@itp.uni-frankfurt.de}

\abstract{The thermodynamics of QCD with sufficiently heavy dynamical quarks can be described by a three-dimensional Polyakov loop effective theory, obtained after a truncated character and hopping expansion. We investigate the resulting phase diagram for low temperatures by mean field methods. Taking into account chemical potentials for 
both baryon number and isospin, we obtain clear signals for a liquid-gas type transition to baryon matter at $\mu_I=0$ and a Bose-Einstein condensation transition at $\mu_B=0$, as well as for their connection when both chemical potentials are non-zero.}

\FullConference{%
The 39th International Symposium on Lattice Field Theory,\\
8th-13th August, 2022,\\
Rheinische Friedrich-Wilhelms-Universität Bonn, Bonn, Germany
}

\begin{document}
\maketitle

\section{Introduction}
The QCD phase structure in the $(T,\mu_B)$ plane is important to understand many physical phenomena for various disciplines in modern physics. The 
interest in neutron stars and their mergers moreover motivates studies of
non-vanishing isospin chemical potential $\mu_I\neq 0$, and in particular the mixed situation with both $\mu_B,\mu_I\neq0$.
While ordinary Monte Carlo simulations are possible for $(\mu_B=0,\mu_I\neq 0)$, cf.~\cite{Brandt:2017oyy} and references therein,
a strong sign problem causes their breakdown whenever the baryon chemical potential $\mu_B \neq 0$. Most methods to circumvent the obstacle introduce further
approximations, and are therefore restricted to the region with $\mu_B/T \leq 3$ (see
for example Refs.~\cite{deForcrand:2009zkb, Aarts:2012yal, Guenther:2020jwe}).

Our approach is to study $3D$ effective Polyakov loop lattice theories, which are
obtained from Wilson's lattice QCD when integrating over all spatial links after a truncated strong coupling and hopping parameter expansion, and describe
QCD with very heavy quarks~\cite{Fromm:2011qi}.
Here we investigate the phase diagram of the simplest of these effective theories
with a mean field approach, and compare with previous solutions obtained by
series expansions or Monte Carlo simulations.
After successful tests with the finite temperature deconfinement transition
we consider $T\rightarrow 0$, where we identify both
the onset transition to cold baryon matter at $\mu_B\approx m_B, \mu_I=0$, 
as well as the transition to a pion condensate at $\mu_B=0,\mu_I=m_\pi/2$. 
Finally, we switch on both chemical potentials and follow the 
critical lines into the $(\mu_B,\mu_I)$-plane. We find them to connect
in a branching point, where the vacuum, baryon and pion phases meet.   

\section{The effective theory}\label{sec_derivation_theory}

In this section we summarise the main features of the effective theory to be analysed below, for details see~\cite{Langelage:2010yr,Fromm:2011qi}. 
Starting point is the path integral
\begin{align}\label{path_integral}
    Z &=  \int \qty[\dd U_{\mu}]~ \Big( \prod^{N_f}_{f=1}
    \det Q_{f}\qty[U_\mu]\Big) e^{-S_{g} \qty[U_{\mu}]}=  \int \qty[\dd U_0] e^{-S_{\text{eff}}[U_0]}, 
\end{align}
with
\begin{align}  \label{eff_action}
     -S_{\text{eff}}[U_0] \equiv \ln \int \qty[\dd U_i] \qty[ \Big( \prod^{N_f}_{f=1}
    \det Q_{f}\qty[U_\mu]\Big) ~ e^{-S_g[U_\mu]}].
\end{align}
An approximation of the effective action Eq.~\eqref{eff_action} can be obtained
after a truncated expansion of the gauge action in terms of the fundamental character coefficients $u(\beta)$ and a hopping expansion
of the fermion determinants in terms of the hopping parameters $\kappa_f$  \cite{Montvay:1994cy},
\begin{equation}
    u(\beta)=\beta/18+\beta^2/216+\ldots,\quad \kappa_f=(2 am_f +8)^{-1},
\end{equation}
followed by analytic integrating over 
the spatial gauge links. Because of the hopping expansion, the resulting
effective theory is only valid for large bare quark masses $m_f$. Here we specialise to
mass-degenerate quarks, for which $m_f=m, \kappa_f=\kappa$ for all flavours $f$.
The remaining dependence on the temporal gauge links is in terms of untraced Wilson lines,
$W(\vb{x}) =\prod^{N_{\tau}}_{\tau = 1} U_{0}\qty(\vb{x}, \tau)$, and Polyakov loops, $L\qty(\vb{x}) = \Tr W(\vb{x})$.
The result after these steps can be written as,
\begin{align}\label{partition_function_final}
    Z &\approx \int DW   \underbrace{\prod_{\expval{\vb{x}, \vb{y}}} \qty[1+\lambda_1 \qty(L^{\dagger}_{\vb{x}} L_{\vb{y}} + L_{\vb{x}} L^{\dagger}_{\vb{y}})]}_{\text{pure gauge}} \\ \nonumber
    &\times \underbrace{\prod_{\vb{x}} \prod^{N_f}_{f=1} \qty( 1 + h_{1f} L_{\vb{x}} + h_{1f}^2 L^{\dagger}_{\vb{x}} + h_{1f}^3)^2 ~ \qty( 1 + \overline{h}_{1f} L^{\dagger}_{\vb{x}} + \overline{h}_{1f}^2 L_{\vb{x}} + \overline{h}_{1f}^3 )^2}_{\text{static}} \\ \nonumber
    &\times \prod_{\expval{\vb{x}, \vb{y}}}\underbrace{\qty(1- 2 h_2\sum^{N_f}_{f=1} \qty[W^f_{11}\qty(\vb{x}) - \overline{W}^f_{11}\qty(\vb{x})]  \qty[W^f_{11}\qty(\vb{y}) - \overline{W}^f_{11}\qty(\vb{y})])}_{\text{kinetic}} 
\end{align}
with the effective couplings \cite{Langelage:2010yr}
\begin{eqnarray}
\lambda_1&=& u^{N_{\tau}} \exp\qty[N_{\tau} \qty(4u^4 + 12u^5 - 14u^6 - 16u^7 + \frac{295}{2} u^8 + \frac{1851}{10} u^9 + \frac{1055797}{5120} u^{10})]  \nonumber \\
h_{1f} &=& 2\kappa\exp\qty[N_\tau a\mu_f] \exp\qty[6N_\tau \kappa^2 u \qty( \frac{1-u^{N_\tau-1}}{1-u} + 4u^4 - 12\kappa^2 + 9\kappa^2 u + 4 \kappa^2 u^2 - 4\kappa^4) ] \nonumber \\ 
\overline{h}_{1f} &=& 2\kappa \exp\qty[-N_\tau a\mu_f] \exp\qty[6N_\tau \kappa^2 u \qty( \frac{1-u^{N_\tau-1}}{1-u} + 4u^4 - 12\kappa^2 + 9\kappa^2 u + 4 \kappa^2 u^2 - 4\kappa^4) ] \nonumber \\ 
h_2&=& \frac{N_{\tau}\kappa^2}{N_c}\qty(1+2\frac{u-u^{N_{\tau}}}{1-u} + 8u^5 + 16\kappa^3 u^4). \nonumber 
\end{eqnarray}
The expressions appearing in the third line in 
Eq.~\eqref{partition_function_final} are defined as 
\begin{align}
    W^f_{nm}\qty(\vb{x}) = \Tr \frac{\qty(h_{1f} W_{\vb{x}})^m}{\qty(\mathbb{I}+h_{1f} W_{\vb{x}})^n}, \quad \overline{W}^f_{nm}\qty(\vb{x}) = \Tr \frac{\qty(\overline{h}_{1f} W^{\dagger}_{\vb{x}})^m}{\qty(\mathbb{I} +\overline{h}_{1f} W^{\dagger}_{\vb{x}})^n}.
\end{align}
The effective theory in Eq.~\eqref{partition_function_final} is structured as follows: 
the first line represents the pure gauge contribution, the second line contains the static determinant, and the last line is the leading order of the kinetic quark determinant. 
All contributions can be expressed fully in terms of Polyakov loops, if desired. 
An important feature of this effective theory is that it has a much weaker sign problem than full QCD and can therefore be simulated using reweighting techniques or 
complex Langevin methods \cite{Fromm:2011qi,Langelage:2014vpa}. Moreover, it can 
also be treated by analytic linked-cluster expansion methods in the 
effective couplings \cite{Glesaaen:2015vtp,Kim:2020atu,Pham:2021ftz}.

\section{Mean field analysis of the effective theory} \label{sec_mean_field}

Since the effective theory resembles a spin model, it is natural to also consider
mean field methods as a short cut to a first evaluation and to get an impression 
of the overall phase structure. We proceed in analogy to early investigations of similar
models \cite{Kogut:1981ez,Green:1983sd}. The basic idea is to 
consider fluctuations of the Polyakov loop around its
mean field value, 
$L_{\vb{x}} = \overline{L} + \delta L_{\vb{x}}$ and $L^{*}_{\vb{x}} = 
\overline{L} + \delta L^{*}_{\vb{x}}$, 
and to expand the effective action up to linear order in the fluctuations, 
\begin{align*}
    S_{\text{eff}}\qty[L] \approx S_{\text{eff}}\qty[\overline{L}] + \sum_{\vb{x}}\qty(\pdv{S_{\text{eff}}}{L}\bigg|_{\overline{L}}\delta L_{\vb{x}} + \pdv{S_{\text{eff}}}{L^{*}}\bigg|_{\overline{L}^{*}}\delta L^{*}_{\vb{x}}) + \dots \;.
\end{align*}
Inserting this mean field approximation, the path integral for the effective 
theory simplifies to
\begin{align} \label{approx_PI}
 Z_\mathrm{mf}=f\qty(\overline{L}) \qty[\int \dd W \exp\qty{ -\pdv{S_{\text{eff}}}{\overline{L}}L -\pdv{S_{\text{eff}}}{\overline{L}^{*}}L^{*}}]^V.
\end{align}
The expression $f\qty(\overline{L})$ represents the saddle point contribution 
to the path integral and reads
\begin{align} \label{FL}
    f\qty(\overline{L}) = \exp\qty[-S_{\text{eff}}[\overline{L}] + \pdv{S_{\text{eff}}}
    {\overline{L}} \overline{L} + \pdv{S_{\text{eff}}}{\overline{L}^{*}} 
    \overline{L}^{*}]\;.   
\end{align}
The remaining integration in Eq.~\eqref{approx_PI} is reduced to one-site integrals,
which can be done after expanding down the exponential, as shown in the example below.
After that we have an explicit formula to calculate the free energy
in the mean field approximation, 
\begin{equation}
    F_\mathrm{mf} = -\ln Z_\mathrm{mf}\;.
\end{equation}

\section{A test case: deconfinement transition in pure gauge theory}

To illustrate our concrete calculations, we list the resulting expressions for the effective theory representing finite temperature Yang-Mills theory, i.e., the partition function  
Eq.~\eqref{partition_function_final} reduced to its first line with one effective coupling only.
This will also provide a test of our mean field procedure.
For the linearised corresponding effective action we have
\begin{align}
    -S_{\text{eff}}[L] \approx  d \sum_{\vb{x}}\qty[\ln\qty(1+2\lambda_1 \abs{\overline{L}}^2) + \frac{2d \lambda_1}{\qty(1+2\lambda_1 \abs{\overline{L}}^2)}
    \qty(\bar{L}^*\delta L_{\vb{x}} +\bar{L}\delta L^*_{\vb{x}})]\;,
\end{align}
where $d=3$ denotes the number of space dimensions.
The saddle point contribution then is
\begin{align}
    f(\overline{L}) = \exp \qty[ Vd\ln \qty(1+2\lambda_1 \abs{\overline{L}}^2 )  + \frac{2Vd}{\qty(1+2\lambda_1 \abs{\overline{L}}^2)}]
\end{align}
and the partition function in the mean field approximation
\begin{align}
    Z_\mathrm{mf} &= f(\overline{L}) \qty[\int \dd W \exp \left\{ \frac{2d \lambda_1 \overline{L}^*}{1+2\lambda_1 \abs{\overline{L}}^2}~L +  \frac{2d \lambda_1 \overline{L}}{1+2\lambda_1 \abs{\overline{L}}^2}~ L^*  \right\} ]^V \nonumber\\
    &=f(\overline{L}) \qty[\sum_{n,m} \frac{\qty(z_1)^n \qty(z_2)^m}{n!m!} 
    \int \dd W ~L^n \qty(L^*)^m ]^V
\end{align} 
with 
\begin{equation}
    z_1 \equiv \frac{2d \lambda_1 \overline{L}^*}{1+2\lambda_1 \abs{\overline{L}}^2}\;,\quad
    z_2 \equiv \frac{2d \lambda_1 \overline{L}}{1+2\lambda_1 \abs{\overline{L}}^2} \;.
\end{equation} 
To proceed we need to evaluate the Polyakov loop integrals, 
for which we we use the formula~\cite{Gattringer:2011gq} 
\begin{align} \label{su3_solver}
    \int \dd W ~L^n \qty(L^*)^m  = \sum^{\lfloor \frac{n}{3} \rfloor}_{j=\max\qty(0, \frac{n-m}{3})} \frac{T(n-m)~2n!~m!~\binom{3 \qty(n-j-\frac{n-m}{3} +1 )}{n-3j} \binom{2j- \frac{n-m}{3} }{j} }{ \qty(n-j-\frac{n-m}{3} +1)! \qty(n-j-\frac{n-m}{3} +2)! 
    \qty(2j-\frac{n-m}{3})!}\;,
\end{align}
with binomial coefficients and the triality function $T(n)=1$ if $n \;\mbox{mod}\; 3=0$, and 
$T(n)=0$ otherwise.

\begin{figure}[t]
    \centering
    \includegraphics[scale=1]{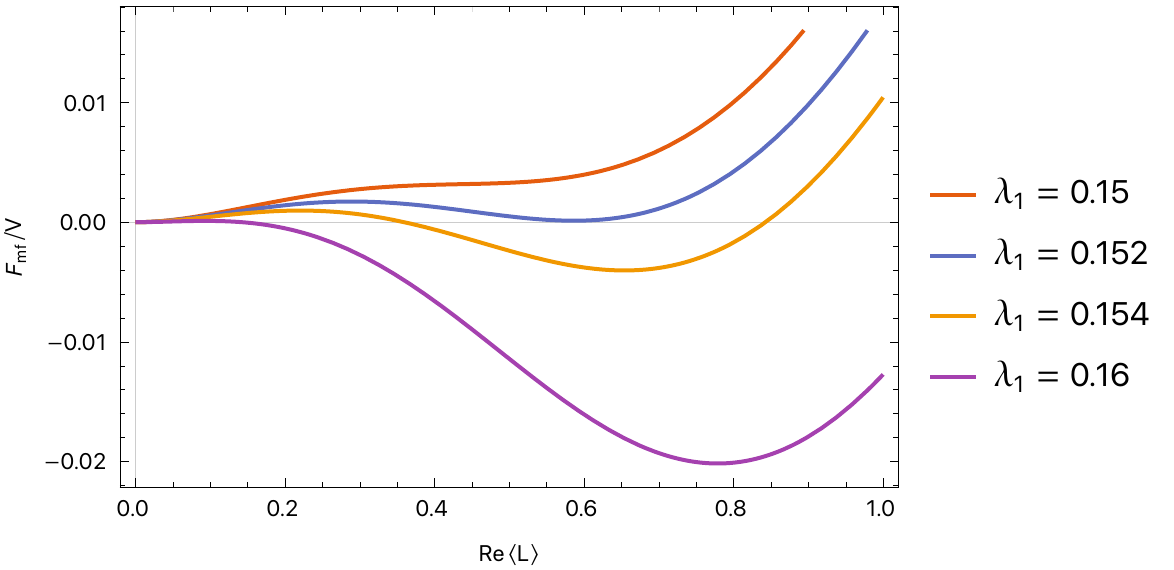}
    \caption{Pure Gauge Case: Mean-Field free energy $\mathcal{F_{\text{mf}}}$ as a function of the expectation value of the Polyalov-Loop $\expval{L}$, with different values around the  critical values of $\lambda_1$.}
    \label{fig:pure_gauge}
\end{figure}

It is well known from lattice simulations that 4d $SU(3)$ pure gauge theory at finite temperature
features a first-order deconfinement transition due to spontaneous 
center symmetry breaking, which is faithfully reproduced by the 3d effective theory \cite{Langelage:2010yr}.
In Fig.~\ref{fig:pure_gauge} we display 
the mean field free energy density as a function of the real part of the Polyakov loop. 
For decreasing effective coupling we indeed observe a non-trivial second minimum to form 
at non-vanishing expectation values, which triggers the center symmetry breaking transition
in the effective theory. Since there is a hill between the minima, this is a first-order 
transition.
The critical coupling for the transition is the value where the two minima are degenerate, 
which happens at $\lambda_{1c}= 0.152$. This is to be compared with 
a Monte Carlo simulation of the 3d effective theory, 
which gives $\lambda_{1c}=0.188$ \cite{Langelage:2010yr}. We conclude that 
the mean field treatment reproduces the correct order of the deconfinement phase transition, and 
the predicted critical coupling is within a reasonable 20\% of the true answer. For an extension of the deconfinement transition to the situation
with dynamical quarks, which is also reproduced, as well as further
refinements of the mean field approach, see \cite{Konrad:2022gwe}.

\section{Phase structure for zero temperature and finite chemical potentials}

After this successful test, we proceed to our case of interest, namely QCD with dynamical
quarks at low temperatures with non-vanishing chemical potentials.
Now Eq.~\eqref{partition_function_final} will be considered with all contribution up to $\kappa^2$ and $N_f=2$. For finite baryon chemical potential, $\mu_u=\mu_d=\mu_B/3$, the 
effective theory has a sign problem, which however is much milder than the original one. It has been demonstrated 
that the effective theory can be simulated with 
a choice of different algorithms. As expected, the theory displays an onset transition to a medium
with net baryon density at $\mu_B\approx m_B$ \cite{Fromm:2012eb,Langelage:2014vpa}.
Like the deconfinement transition, our mean field treatment reproduces 
this baryon onset to be first-order for 
sufficiently large $N_\tau$, i.e.~low temperatures, by
our mean field treatment.

\begin{figure}[t]
    \centering
    \includegraphics[scale=1]{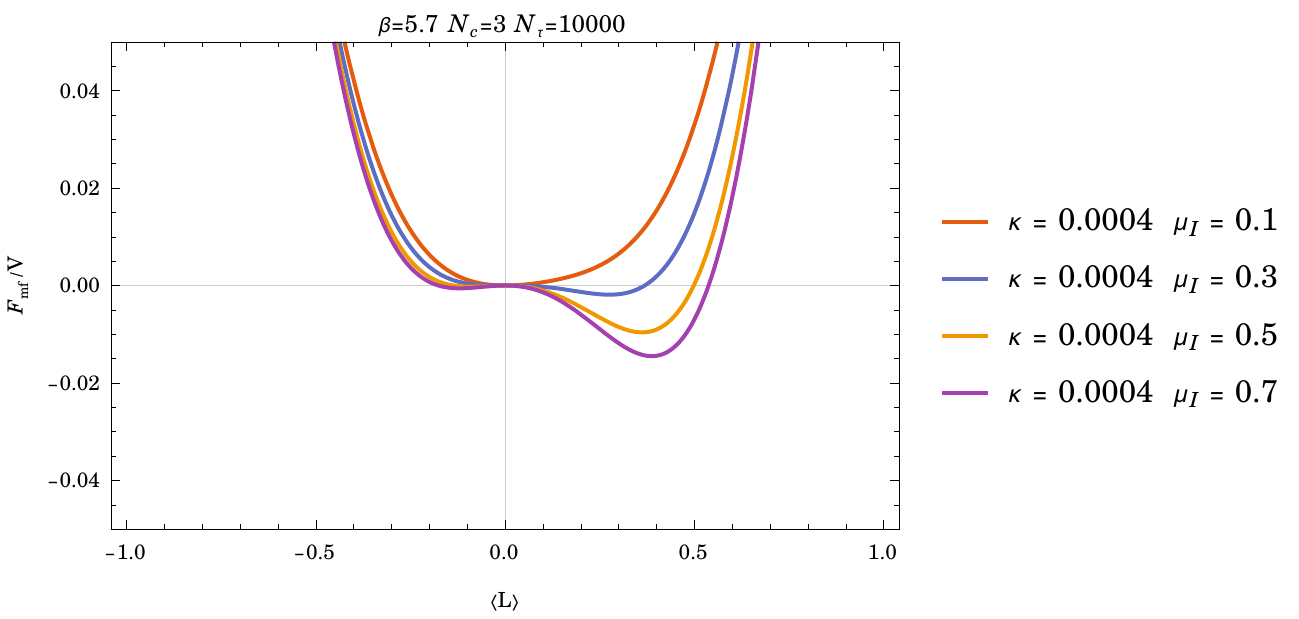}
    \caption{Variation of the mean field free energy density with the isospin chemical potential,  lattice parameters: $\beta=5.7$ and $N_{\tau} = 10000$}
    \label{fig:iso}
\end{figure}

Here, we investigate for the first time how the effective theory behaves when isospin
chemical potenital is introduced, $\mu_I = \qty(\mu_u - \mu_d)/2$.  
In Fig.~\ref{fig:iso} we display the mean field free energy density for the QCD lattice 
parameters  $\beta = 5.7, \kappa=0.0004$ and $N_{\tau} = 10000$. Upon increasing isospin chemical
potential, we observe a flattening of the potential well and the formation of a minimum for 
a non-vanishing value of the Polyakov loop expectation value.
In contrast to the previous transitions, there is no hill in the free energy density separating minima,
but instead a critical value of the chemical potential where the curvature vanishes.
This feature predicts the phase transition to be second order. Once again, this result is fully 
compatible with what is found in simulations of full QCD at the physical point 
with an isospin chemical potential~\cite{Brandt:2017oyy}.

\begin{figure}[t]
    \centering
    \includegraphics[scale=0.45]{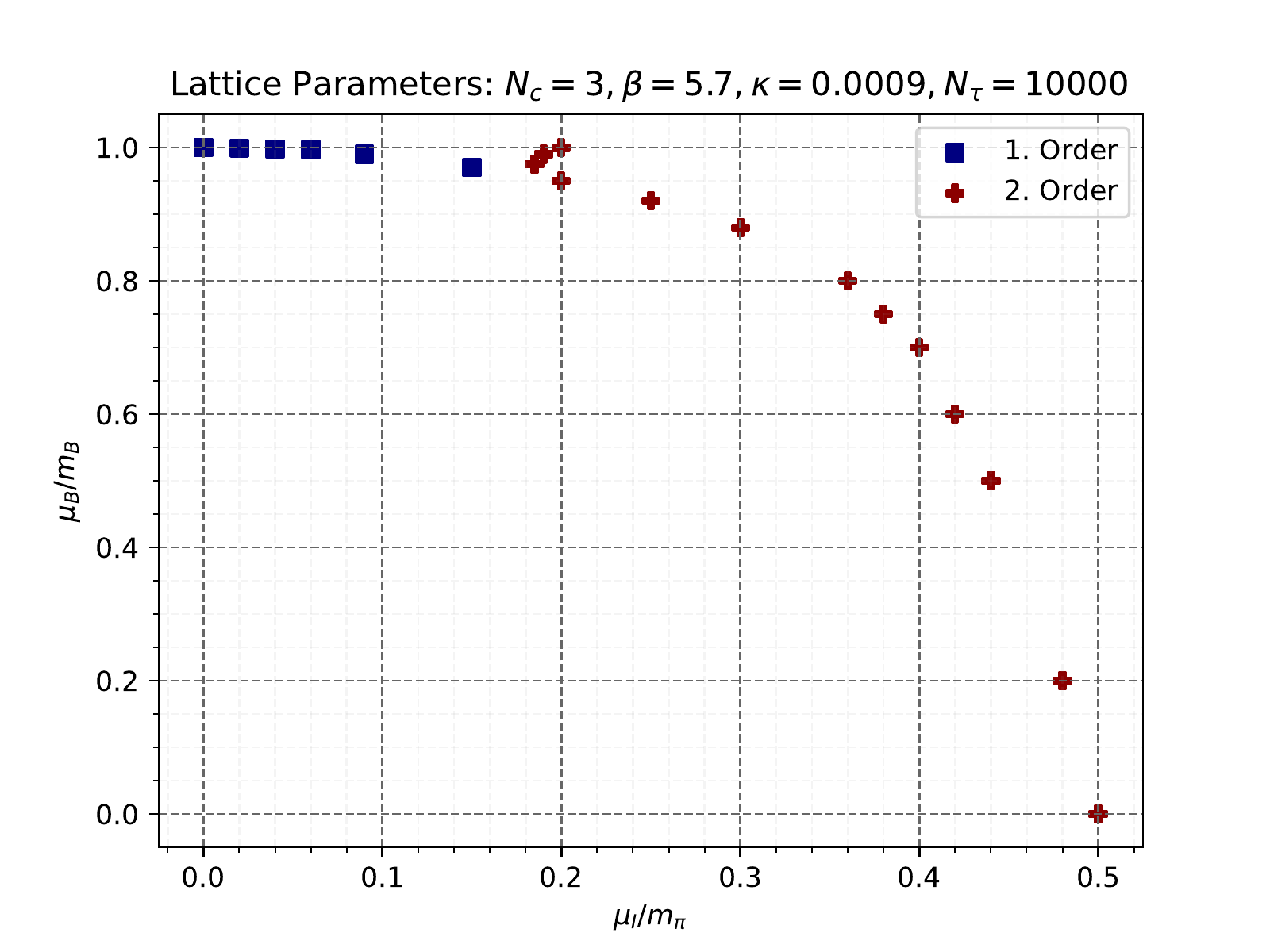}
    \caption{$\qty(\mu_I, \mu_B)$-Phase diagram for $T\rightarrow 0$.}
    \label{fig:phase_diagram}
\end{figure}

In contrast to any other method, our effective theory now permits us to consider
non-vanishing baryon and isospin chemical potential at the same time, and to follow
what happens to the observed first-order baryon onset and second-order isospin condensation 
transitions, respectively. That is, we keep $N_\tau=10 000$ fixed and identify the critical 
combinations $(\mu_B^c,\mu_I^c)$ by either coexistence of two degenerate minima (first order)
or a vanishing curvature (second order) in the mean field free energy density.
The result is shown in Fig.~\ref{fig:phase_diagram}, where the chemical potentials are
given in terms of the baryon and pion masses, which are here
crudely approximated by their leading order values in the hopping expansion, 
$am_B=-3\ln(2\kappa),\; am_\pi=-2\ln(2\kappa)$.

In Fig.~\ref{fig:phase_diagram} we observe both transitions to curve towards each other,
and to join in a special point, where the order of the transition changes.
That the baryon and pion onset transitions must connect somewhere 
is also to be expected before an explicit calculation. The region below the
transition line represents vacuum, whereas above it the ground state of the system 
is either baryon matter or a Bose-Einstein condensate of pions. 
In both of the latter cases there is a medium whose rest frame  
breaks Lorentz symmetry. Thus a true phase transition must exist, independent
of the direction in the plane of chemical potentials. Similarly,
a Bose-Einstein condensate $\langle m_\pi^\pm\rangle\neq0$ and a non-vanishing baryon expectation value 
$\langle n_B\rangle\neq 0$ are distinguished by quantum numbers pertaining to different symmetries.
Consequently, baryon matter must be separated from pion matter by a true phase transition.
Indeed, we observe a second-order transition line separating those phases to emanate from 
the meeting point of the two transition lines from vacuum to matter. Unfortunately, 
we cannot yet follow this line to larger chemical potentials, which is precluded by lattice saturation.

\section{Conclusions}

In this work, we have studied three-dimensional effective lattice theories
for heavy quark QCD via a mean field approximation. This approach is promising because the effective theories represent $SU(3)$ spin models, and one expects
at least qualitatively correct results for the phase structure of the
effective theories. Indeed, for the simplest case representing the 
finite temperature pure gauge theory, a first-order deconfinement 
transition is observed, and the critical effective 
coupling is reproduced within $\sim$20\% of the true answer.

We have then studied the situation close to the zero temperature limit,
when both a baryon chemical potential and an isospin chemical potential
are switched on. We obtain a clear picture with a first-order baryon
onset transition for $\mu_B\approx m_B, \mu_I=0$ and a second-order 
transition to a phase with isospin or
Bose-Einstein condensate at $\mu_I\approx m_\pi/2, \mu_B=0$. 
These transitions are continuously connected and separate vacuum from 
matter. They meet in a branch point,
where an additional second-order line emerges to separate the baryon region
from the pion condensate region.

\acknowledgments
The authors acknowledge support by the 
Deutsche Forschungsgemeinschaft (DFG) through the 
grant CRC-TR 211 ``Strong-interaction matter under extreme conditions''
and
 by the State of Hesse within the Research Cluster ELEMENTS (Project ID 500/10.006).

\bibliographystyle{JHEP}           
\bibliography{biblio}

\end{document}